\documentstyle[preprint,eqsecnum,aps,epsf]{revtex}
\newif\iftightenlines\tightenlinesfalse
\tightenlines\tightenlinestrue
\def\eslt{\not\!\!{E_T}}
\def\etmiss{\not\!\!{E_T}}
\def\to{\rightarrow}
\def\tb{\bar{t}}
\def\bb{\bar{b}}
\begin{document}
%
%
\preprint{\vbox{\baselineskip=14pt
   \rightline{FSU-HEP-990630}\break
   \rightline{BNL-HET-99-14}\break
}} 

%
%
\title{MEASURING THE TOP QUARK YUKAWA COUPLING\\
AT A LINEAR $e^+e^-$ COLLIDER}
\author{Howard Baer$^1$, Sally Dawson$^2$ and Laura Reina$^1$}
\address{
$^1$Physics Department,
Florida State University,
Tallahassee, FL 32306, USA
}
\address{
$^2$Physics Department,
Brookhaven National Laboratory,
Upton, NY 11973, USA
}
\date{\today}
\maketitle
\begin{abstract}
  
  The cross section for the reaction $e^+e^- \to t\tb H$ depends
  sensitively on the top quark Yukwawa coupling $\lambda_t$. We
  calculate the rate for $t\bar{t}H$ production, followed by the decay
  $H\to b\bar{b}$, for a Standard Model Higgs boson with
  $100\,\mbox{GeV}\!\le\! m_H\!\le\!130\,\mbox{GeV}$. We interface
  with ISAJET to generate QCD radiation, hadronization and particle
  decays.  We also calculate the dominant $t\tb b\bb$ backgrounds from
  electroweak and QCD processes.  We consider both semileptonic and
  fully hadronic decays of the $t\tb$ system.  In our analysis, we
  attempt full reconstruction of the top quark and $W$ boson masses in
  the generated events.  The invariant mass of the remaining $b$-jets
  should show evidence of Higgs boson production.  We estimate the
  accuracy with which $\lambda_t$ can be measured at a linear $e^+e^-$
  collider. Our results, including statistical but not systematic
  errors, show that the top quark Yukawa coupling can be measured to
  6--8\% accuracy with 1000 fb$^{-1}$ at $E_{CM}=1$~TeV, assuming
  100\% efficiency for $b$-jet tagging. The accuracy of the
  measurement drops to 17--22\% if only a 60\% efficiency for
  $b$-tagging is achieved.

\end{abstract}

\medskip



\section{Introduction}
\label{sec:intro}
\def\beqn{\begin{eqnarray}}
\def\eeqn{\end{eqnarray}}
\def\be{\begin{equation}}
\def\ee{\end{equation}}

One of the most important challenges facing the next generation of
accelerators is the untangling of the mechanism of electroweak
symmetry breaking.  In the optimistic scenario where a Higgs boson is
discovered at LEP2, the Tevatron, or the LHC, a major goal of a high
energy $e^+e^-$ collider becomes the measurement of the Higgs boson
couplings to the fermions and gauge bosons.  These couplings are
uniquely predicted in the Standard Model (SM), but can be
significantly different in extentions such as supersymmetric models.
A measurement of the Higgs couplings can therefore discriminate
between various alternatives to the Standard Model.
  
The couplings of the Higgs boson to the gauge bosons can be measured
in a straightforward manner through the associated production
processes, $e^+e^-\rightarrow Z H$, $q {\overline
q}^\prime\rightarrow W^\pm H $, and $q {\overline q}\rightarrow Z H $,
and through vector boson fusion, $W^+W^-\rightarrow\! H $ and
$ZZ\rightarrow H $.  The couplings of the Higgs boson to fermions are
more difficult to measure, however.  In the Standard Model, the
fermion-Higgs boson couplings are completely determined in terms of
the fermion masses.  For a generic quark, $Q$, the $Q {\overline Q} H$
coupling is given by,

\be \lambda_Q=-{M_Q\over v}\,\,\,,
\label{yuk}
\ee 

\noindent 
where $v\!=\!(\sqrt{2}G_F)^{-1/2}$ and so the top quark-Higgs boson
Yukawa coupling is much larger than any other Yukawa coupling.
  
The process $e^+e^-\rightarrow t {\overline t} H$ provides a direct
mechanism for measuring the top quark-Higgs Yukawa
coupling\cite{zer,desystud}. This process proceeds mainly via $\gamma$
and $Z$ exchange, with the Higgs boson radiated from the top quark.
This contribution is directly proportional to $\lambda_t^2$.  There is
also a contribution from the Higgs boson coupling to the exchanged
$Z$, but this contribution is subdominant and so does not upset the
interpretation of the $e^+e^- \to t \tb H$ process as a measurement of
the top quark-Higgs boson Yukawa coupling.

At a high energy $e^+e^-$ collider with $E_{CM}=500$~GeV, the process
$e^+e^- \to t \tb H$ is sensitive to Higgs bosons in the mass region
$100~\mbox{GeV}\sim m_H\sim 130~\mbox{GeV}$.  The current limit on
$m_H$ from LEP2 is $m_H>95.2$~GeV\cite{lep2higgs}.  This Higgs mass
regime is favored by calculations comparing precision electroweak data
to Standard Model predictions\cite{erler}.  In addition, the lightest
Higgs boson of supersymmetric models (which for many models behaves
indistinguishably from the SM Higgs boson) ought to have mass less
than typically 130~GeV\cite{susyhmass}.  In the interesting
$100\!-\!130~\mbox{GeV}$ mass region, the Higgs boson will decay
predominantly to $b \bb$ pairs and so the final state of interest will
be $t \tb b \bb$.  Since the top quark decays to a $W$ boson and a $b$
quark, the final state will contain at least four $b$-quarks plus two
$W$ bosons.  Although the rate is small (on the order of a few
femtobarns), the signature is quite spectacular.

There are two major sources of background to the $t \tb b \bb$ final
state\cite{moretti}.  There is a large QCD background, coming
primarily from the process $e^+e^- \to t \tb g^*$, with the gluon
decaying to a $b \bb$ pair. The $b \bb$ pairs resulting from the gluon
splitting tend to peak at low values of the $b \bb$ invariant mass.
There is also an electroweak (EW) background, of which the dominant
contribution is $e^+e^-\to Z t \tb$. Although the EW background is
formally smaller than the QCD background, it is more problematic since
the $Z\to b\bb$ decay resonates in close proximity to the expected
Higgs signal.  We compute these backgrounds and estimate the resulting
precision which can be obtained on $\lambda_t$.

The $O(\alpha_s)$ cross section for $e^+e^-\rightarrow t\bar t H$ has
been calculated in Refs.~\cite{sdlr,spiraetal}.  In this paper, we
work entirely with tree level cross sections, for consistency with the
background predictions.  The dominant effect of QCD corrections will
be to increase the precision of signal and background total cross
section calculations, so that results from this paper can be
appropriately rescaled once the QCD corrected background rates are
available.

Recently, Moretti has presented parton-level calculations of both
signal and background processes for the semileptonic final
state~\cite{moretti}.  We have confirmed both his signal and
background results at the parton level.  In this paper, we extend our
programs to include parton showering, hadronization and particle
decays.  We consider both the semileptonic final state

\begin{eqnarray*}
e^+e^-\to (b\ell\nu )+(bq\bar{q}')+(H\to b\bb ) \,\,\,,
\end{eqnarray*}
and the fully hadronic final state
\begin{eqnarray*}
e^+e^-\to (bq_1\bar{q_1}')+(bq_2\bar{q}_2')+(H\to b\bb )\,\,\, .
\end{eqnarray*}
In our approach, we rely on a full reconstruction of the various top
quark, $W$-boson and Higgs boson invariant masses in the events, which
should lend confidence that the appropriate signal and background
processes are in fact being seen.

Similar analyses of top quark Yukawa coupling measurements have
recently been presented at conferences.  A recent analysis of signal
and background for $t\bar t H$ production and decay for both
semileptonic and hadronic channels has been presented by Juste and
Merino using a neural net approach~\cite{merino}. An independent
analysis is also being carried out by the authors of
Ref.~\cite{schreiber}.

The associated $e^+e^-\to Q {\overline Q} H_i$ process (for
$Q\!=\!t,b$) is also of interest in supersymmetric
models\cite{zer,susy,grz}.  In such models, there are five Higgs
bosons, $H_i$, which can be produced and the couplings of the Higgs
bosons may differ significantly from those of the Standard Model.  In
particular, the coupling of at least one of the Higgs bosons to the
$b$ quark may be enhanced.  In addition, because of the tri-linear
coupling of the Higgs bosons to a scalar and a pseudoscalar, in such
models the $b\bb H$ production receives large resonance contributions
not present in the Standard Model.  These models offer a rich
phenomenology, but will not be considered here.

In this report, Section \ref{sec:calculation} contains a description
of some of our calculational details. Section \ref{sec:results_500}
presents results for a linear $e^+e^-$ collider operating at
$E_{CM}=500~\mbox{GeV}$, while Section \ref{sec:results_1000} shows
corresponding results for $E_{CM}\!=\!1$~TeV.  In Section
\ref{sec:conclusions}, we present a discussion and some conclusions.

\section{Calculational Details}
\label{sec:calculation}

The starting point for our signal and background calculations is the
calculation of the corresponding squared matrix elements for the
relevant subprocesses.  For these, we use the computer program
MADGRAPH\cite{madgraph} and the HELAS subroutines\cite{helas}.  We
compute :\footnote{Graphs involving the Higgs boson have been
removed from the EW background to avoid double counting.}
\begin{eqnarray*}
e^+e^-&\to & t\tb H \ \ \ \ (5\ {\rm diagrams})\,\,, \\
e^+e^-&\to & t\tb b\bb \ \ \ \ (EW:\ 35\ {\rm diagrams})\,\,,\ {\rm and} \\
e^+e^-&\to & t\tb b\bb \ \ \ \ (QCD:\ 8\ {\rm diagrams})\,\,.
\end{eqnarray*}

In the QCD contribution, we have taken $\alpha_s(M_Z)=.118$.  We
notice that, with respect to Ref.~\cite{merino}, our background
calculation includes the Z-boson spin correlation effects for the
electroweak background. Moreover the dominant QCD background is fully
simulated.

The squared matrix elements are integrated via Monte Carlo over phase
space, and the total cross sections and distributions agree with the
results of Moretti\cite{moretti}. Next, we interface our parton-level
programs with the event generator ISAJET\cite{isajet} to allow for
parton showers, hadronization and particle decays. We neglect initial
state bremsstrahlung and beamstrahlung effects. For our analysis,
these effects should mainly lead to a small rescaling of both the
signal and background production cross sections. We also neglect the
spin correlation of the top quark between production and decay, but
expect this to be a small effect as well.

In our analysis, we use the ISAJET toy detector simulation program
ISAPLT. We assume calorimetry in the range $-4 < \eta < 4$,
with cell size $\Delta\eta\times\Delta\phi =0.1\times 0.262$.  We take
the electromagnetic energy resolution to be $0.15/\sqrt{E}\oplus 0.01$
and the hadronic calorimeter resolution to be $0.5/\sqrt{E}\oplus
0.02$ ($E$ in GeV). Calorimeter cells are coalesced in towers of
$\Delta R=\sqrt{(\Delta\eta )^2+(\Delta\phi )^2} =0.5$ using the jet
finding algorithm GETJET.  Hadronic clusters with $E_T(j)>15$~GeV are
called jets.  Leptons ($e$'s or $\mu$'s) with $p_T$ of 5~GeV or more are
considered to be isolated if the hadronic $E_T$ in a cone about the
lepton of $\Delta R=0.4$ is less than 2~GeV.  Jets are classified as
$b$-jets with a tag efficiency of $\epsilon_b$ if they coincide with
an original $b$-parton within an angle $\Delta R=0.4\,$.

\section{Observability at $E_{CM}=500$~GeV}
\label{sec:results_500}
\subsection{Semileptonic channel}
\label{subsec:slchannel_500}

To examine events in the semileptonic channel, we require :
\begin{itemize}
\item one and only one isolated $e$ or $\mu$ with $E>15$~GeV,
\item $\etmiss >15$~GeV,
\item exactly four tagged $b$-jets,
\item $\ge 2$ non-$b$-jets,
\item $60<m$(non-$b$-jets)$<90$~GeV (consistent with $M_W$).
\end{itemize}

The number of events expected for 1000 fb$^{-1}$ of integrated
luminosity at $E_{CM}=500$~GeV is shown before and after these cuts in
the first two rows of Table~\ref{tab:eve_l_500}. For the time being,
we take $\epsilon_b =1$.  Already at this stage we see a huge
reduction in the QCD background. The major QCD background reduction
comes from the requirement of $\ge 4$ $b$-jets in the final state: for
this background, the $b$'s from $g^*\to b\bb$ are usually relatively
soft and not well separated in angle, so only rarely do we get four
distinct $b$-jets from this process.  The signal is already well in
excess of background for $m_H=100$ and 110~GeV, and ranges from $3-33$
events for the $m_H$ values we have chosen.  An integrated luminosity
of order 1000 fb$^{-1}$ will be essential for this measurement at
$E_{CM}=500$~GeV.

At this stage, we can attempt to reconstruct some of the invariant
masses that occur in these events. As an example, we show various mass
distributions in Fig.~\ref{fig:masses}. These distributions were
generated for $m_H=120$~GeV and $E_{CM}=1$~TeV; the results for
$E_{CM}=500$~GeV are qualitatively similar.  First, in
Fig.~\ref{fig:masses}{\it a}, the invariant mass of {\it all}
non-$b$-jets in the events (before imposing the 60-90~GeV cut listed
above) is shown. The invariant mass rises to a peak near to, but
slightly below, $m(\mbox{non}\!-\!b\!-\!\mbox{jets})=M_W$. The peak occurs
below $M_W$ mainly due to jet activity leaking out of our fixed cone
algorithm, so better jet reconstruction algorithms may improve upon
this.  We impose the $60<m$(non-$b$-jets)$<90$~GeV cut to insure
events consistent with a hadronic $W$-boson decay.

In Fig.~\ref{fig:masses}{\it b}, we combine the non-$b$-jets with the
$b$-jet which most nearly reconstructs the top quark mass. The
distribution peaks sharply just below $m_t$, due in part to missing
neutrinos from $B$-meson cascade decays, along with leakage from the
jet cones.  In Fig.~\ref{fig:masses}{\it c}, we attempt to reconstruct
the other $W\to\ell\nu$ decay. Beam- and bremsstrahlung effects do not
allow us to use the beam-beam center-of-mass energy to constrain the
$z$ component of missing energy, so we work instead with transverse
mass.  The transverse mass distribution is shown, and peaks as
expected just below $m_T(\ell,\eslt )=M_W$, with considerable smearing
due in part to $B$ and $D$ meson semileptonic decay
contamination. Finally, in Fig.~\ref{fig:masses}{\it d}, we
reconstruct the $b\,\ell +\etmiss$ cluster transverse mass. We pick one
of the remaining $b$-jets which most nearly reconstructs to $\le
m_t$. The peak from the top quark decay is again visible.  If our mass
reconstruction algorithm has been successful, then the remaining two
$b$-jets should reconstruct to $m_H$ for our signal events, but to
other values for background events.

At this point, we note that the energy distribution of $b$-jets should
vary considerably between signal and background. This should
especially hold true for the slowest (least energetic) of the
$b$-jets. We plot in Fig.~\ref{fig:ebsl} the energy distribution of
the slowest of $b$-jets, $E_b(slow)$, at $E_{CM}=500$~GeV, after the
above cuts. The dashed histogram for the sum of all background
processes peaks at low $E_b(slow)$, while the $E_b(slow)$ distribution
for the signal becomes increasingly harder for heavier Higgs boson
masses. To gain some improvement in signal-to-background ratio (S/B),
we will impose
\begin{itemize}
\item $E_b(slow)>25,\ 40\ {\rm and}\  45\ {\rm  GeV}$
\end{itemize}
for $m_H=110,\ 120$ and 130~GeV.

After the above cuts and mass reconstructions, we plot in
Fig.~\ref{fig:xmbbl5} the invariant mass of the remaining two
$b$-jets.  The signal histograms are solid, while background is
dashed. For the $m_H\!=\!100$~GeV case, the mass distribution peaks
somewhat below 100~GeV with a rather broad smear. In this case, the
distribution in energy for $b$-jets from Higgs decay is similar to the
energy distribution of $b$-jets from top decay, so our reconstruction
algorithm often fails to select the correct $b$-jets from Higgs
decay. In addition, neutrinos from $B$ and $D$ meson decay serve to
further soften the distribution. As we increase the Higgs mass in
frames {\it b})--{\it d}), the signal distribution becomes harder and
sharper.

For comparison, we show in Fig.~\ref{fig:xmbbexl5} the $m(bb)$
distribution using generator information to select the correct
$b$-quark jets. These distributions may be approached, for instance,
by using additional information such as the previously measured Higgs
mass in the reconstruction algorithm. For these cases, the Higgs mass
peak is more clearly defined, as is the peak in the background from
the $Z$ resonance.

Our results for this channel at $E_{CM}\!=\!500$~GeV are collected in
Tables~\ref{tab:eve_l_500} and \ref{tab:err_l_500}.  The columns
labelled with the Higgs boson mass ($H(100)$, {\it etc}) are the
signal for $e^+e^-\to t \tb H$, $H\to b\bb$ followed by decays
yielding four $b$-quark jets plus a lepton plus missing energy.  The
columns labelled $ t \tb b \bb(EW)$ and $ t \tb b \bb(QCD)$ give the
electroweak and QCD backgrounds respectively.

The first row of Table~\ref{tab:eve_l_500} gives the total number of
$t \tb b \bb$ events before the top quark decays. In the second row we
report the number of events reconstructed for the specific channel.
By placing cuts on the $m(bb)$ mass distribution, some further
improvement in S/B can be gained. We list in the following rows of
Table~\ref{tab:eve_l_500} the events expected after selected cuts on
$E_b(slow)$ and $m(bb)$.  After cuts, the remaining number of events
is between $1$ and $34$ for $1000$~fb$^{-1}$.  Even with this
optimistically high luminosity, there is less than one remaining
signal event for $m_H=130~\mbox{GeV}$.

In Table~\ref{tab:err_l_500}, we show the resulting statistical
precision for the measurement of $\delta \lambda_t/\lambda_t$, for
$\epsilon_b =1$ and $0.6$.  The error is calculated assuming a Poisson
distribution for both signal and background, and is given by
\be
\frac{\delta \lambda_t}{\lambda_t} = \frac{1}{\sqrt{S}}\sqrt{1+\frac{2
      B}{S}}\,\,\,,
\label{poisson}
\ee
where $S$ and $B$ are respectively the number of signal and background
events.  We have assumed that $\Gamma(H\to b\bb)$ will be known
precisely from previous measurements at the LHC and NLC, and so the
signal rate for $e^+ e^-\to t \tb H,~ H\to b\bb$ depends only on
$\lambda_t$.  If high efficiency can be achieved on $b$-jet tagging,
then already at $E_{CM}=500$~GeV the top quark Yukawa coupling can be
measured to 11\% for $m_H=100$~GeV. This decreases to 31\% if only
$\epsilon_b=0.6$ is achieved.  The precision
${\delta\lambda_t\over\lambda_t}$ becomes rapidly worse as $m_H$
increases beyond $100~\mbox{GeV}$.  For $m_H=120~\mbox{GeV}$ and an
integrated luminosity $L$, we find,
\be {\delta \lambda_t\over\lambda_t} \sim 39\%\sqrt{{ 1000\,
    \mbox{fb}^{-1}\over L \epsilon_b^4}}\,\,\,, \quad\quad
    m_H=120~\mbox{GeV}\,\,\,.
\ee 
It appears that $E_{CM}=500~\mbox{GeV}$ is a poor energy for the
measurement of the $t \tb H$ Yukawa coupling unless the Higgs boson is
$\sim 100$~GeV.  At this energy, the $t\tb H$ system is close to the
phase space limit and the parent particles have little kinetic energy,
making the kinematic cuts and event reconstruction less effective than
at higher energies.  In addition, the number of events is too small to
obtain a statistically precise measurement for the heavier Higgs boson
masses.

\subsection{Hadronic channel}
\label{subsec:hchannel_500}

The totally hadronic channel for $e^+e^-\to t\tb H$ has the advantage
of initially higher rates than the semileptonic channel due to the
large $W$ boson hadronic branching fraction. However, in attempting
mass reconstructions, a greater combinatoric problem is presented
since there will now be typically four or more non-$b$-jets in each
event, in addition to the four $b$-jets.

For the hadronic channel, we make the following cuts :
\begin{itemize}
\item exactly zero isolated leptons with $p_T>5$~GeV,
\item exactly four identified $b$-jets,
\item $\ge 4$ non-$b$-jets .
\end{itemize}
We then attempt mass reconstruction. In Fig.~\ref{fig:massh}{\it a},
we show the invariant mass of the two non-$b$-jets $m_1(jj)$ that most
nearly reconstructs $M_W$. Again, for illustration, we show these
results for $m_H=120$~GeV and $E_{CM}=1$~TeV; results for
$E_{CM}=500$~GeV are qualitatively similar. The resonance from the $W$
boson is evident. In Fig.~\ref{fig:massh}{\it b}, we cluster the two
jets from {\it a}) with the $b$-jet which most nearly reconstructs
$m_t$; the distribution has the expected peak just below
$m_1(bjj)=m_t$.  In Fig.~\ref{fig:massh}{\it c}, we plot the invariant
mass of {\it all} the remaining non-$b$-jets. In this case, we again
have a peak near $m_2(jets)=M_W$, but with significant smearing below
and above the peak.  Likewise, in Fig.~\ref{fig:massh}{\it d} we show
the invariant mass of the jets from {\it c}) with the remaining
$b$-jet which most nearly reconstructs $m_t$. The distribution peaks
at $m_t$, but again with significant smearing, especially above
$m_2(b,jets)=m_t$.  To be consistent with reconstructing a second
hadronic $W$ and a second hadronic top quark, we require only events
with
\begin{itemize}
\item $60<m_2(jets)<90$~GeV , and
\item $125<m_2(b,jets)<200$~GeV.
\end{itemize}
The resulting event rates are listed in Table~\ref{tab:eve_h_500},
following the same pattern explained in
Sec.~\ref{subsec:slchannel_500}.  After these additional cuts, the
surviving number of events in 1000 fb$^{-1}$ is surprisingly close to
the number of events expected in the semileptonic channel.

At this point, we may apply the same $b$-jet energy cuts as in the
semileptonic case, and plot the mass of the remaining $b$-jets in the
events. These results are shown in Figs.~\ref{fig:xmbbh5} and
\ref{fig:xmbbexh5} for the remaining $b$-jets and the exact
reconstruction, respectively.  As before, some improvement in S/B can
be made by adopting a $m(bb)$ mass cut. These are listed in
Table~\ref{tab:eve_h_500}, along with the surviving number of
events. In a similar fashion to the semileptonic case, we can then
extract the error measurements on the top quark Yukawa coupling, and
these are listed in Table~\ref{tab:err_h_500} for $\epsilon_b=1$ and
$0.6$. In the hadronic channel, the results are very similar to the
leptonic case, and so will offer an independent confirmation of any
sort of top quark Yukawa measurement.  Of course, the semileptonic and
hadronic channel measurements can be combined to improve the overall
precision of the measurement.

\section{Observability at $E_{CM}=1$~TeV}
\label{sec:results_1000}

\subsection{Semileptonic channel}

At $E_{CM}=1$~TeV, the total event rates are significantly larger than
at $E_{CM}=500$~GeV.  Since this energy scale is far above the
kinematic limit, there is only a modest sensitivity of the total
signal rate to the Higgs boson mass.  At $E_{CM}=1$~TeV, the energy
distribution of the slowest $b$ jet is not as distinctive as it was in
the $E_{CM}=500$~GeV, due mainly to the high momentum of the parent
particles that are produced. Hence, we drop the $E_b(slow)$ cut for this
energy regime.  We do adopt the remaining semileptonic cuts from
Section \ref{sec:results_500} and show the resulting signal and
background rates in Table~\ref{tab:eve_l_1000}. After selection cuts,
about 60 background events should remain while 60-110 signal events
would be present, depending on the value of $m_H$. We apply the same
mass reconstruction algorithm as in Sec.~\ref{sec:results_500}, and
plot the invariant mass of the remaining $bb$ pair in
Fig.~\ref{fig:xmbbl1}, and the mass of the correctly identified $bb$
pair in Fig.~\ref{fig:xmbbexl1}. The mass
reconstruction in Fig.~\ref{fig:xmbbl1} is far sharper than the
corresponding plot at 500~GeV. In this case, the large kinetic energy
of the parents is transferred to the daughter particles, and wrong
mass reconstructions become much more difficult. We may again apply a
$m(bb)$ mass cut, as listed in Table ~\ref{tab:eve_l_1000}, to
improve the S/B ratio. The corresponding precision on the top quark
Yukawa coupling measurement is listed in Table~\ref{tab:err_l_1000}.
Although the contribution of the $ZZH$-vertex diagram increases at
$E_{CM}\!=\!1$~TeV with respect to $E_{CM}\!=\!500$~GeV, it is still
of the order of a few percent and completely negligible in the
determination of the error on $\lambda_t$.

From Table~\ref{tab:err_l_1000}, we see that there will be roughly a
$7-9\%$ measurement of $\delta\lambda_t/\lambda_t$ in the leptonic
channels with $1000$~fb$^{-1}$ for $\epsilon_b=1$.  These results are
degraded to $19-24\%$ if only $\epsilon_b=0.6$ can be achieved.  For
the measurement of the top quark Yukawa coupling, $E_{CM}=1$~TeV is
clearly far superior to $E_{CM}=500$~GeV for $m_H>100$~GeV.

\subsection{Hadronic channel}

Finally, we present results for the hadronic channel at
$E_{CM}=1$~TeV.  For this case, we apply again the same cuts as in
Sec.~\ref{sec:results_500}, but without the cut on $E_b(slow)$. The results
are given in Table~\ref{tab:eve_h_1000}. The reconstructed $m(bb)$
and exact $m(bb)$ are shown in Figs~\ref{fig:xmbbh1} and
\ref{fig:xmbbexh1}.  The corresponding results after applying a cut on
$m(bb)$ are again listed in Table ~\ref{tab:eve_h_1000} and range
from 20-50 events for 1000 fb$^{-1}$ of data. The precision on the top
quark Yukawa coupling is given in Table~\ref{tab:err_h_1000}. The
precision ranges from $10-14\%$ for $\epsilon_b=1$, and from $28-39\%$
if only $\epsilon_b=0.6$ is achieved.

\section{Conclusions}
\label{sec:conclusions}

The process $e^+e^-\to t \tb H$ directly measures the $t \tb H$ Yukawa
coupling.  We have computed the signal and the major backgrounds for
both the semileptonic and hadronic decay channels using ISAJET to
simulate gluon radiation, hadronization and decays.  In our analysis,
we rely on a direct reconstruction of the $W$, $t$ and $H$ masses in
the events. At $E_{CM}=500$~GeV, even with 1000 fb$^{-1}$ of
integrated luminosity, only $m_H\alt 110$~GeV will give enough event
rate for a reasonable measurement of the top quark Yukawa coupling.
At higher energies, the entire range of $m_H=100-130$~GeV should be
accessible.

Our final results indicate the statistical error that can be achieved
on the measurement of the top quark Yukawa coupling. Systematic errors
will also be present, but these will depend in detail on properties of
the machine and detector, so we do not attempt to estimate these.  Our
final results can be quoted by combining the best measurement in the
semileptonic channel with the best measurement in the hadronic
channel, as two independent measurements. We then find for
$\delta\lambda_t /\lambda_t$ at $E_{CM}=500$~GeV and 1000 fb$^{-1}$,

\begin{center}
\begin{tabular}{ccc}
$m_H(\mbox{GeV})$ & \hspace{.5truecm}$\epsilon_b\!=\!1$ & 
\hspace{.5truecm}$\epsilon_b\!=\!0.6$ \\
100 &\hspace{.5truecm} 0.08 &\hspace{.5truecm} 0.22 \\
110 &\hspace{.5truecm} 0.12 &\hspace{.5truecm} 0.32 \\
120 & \hspace{.5truecm}0.21 & \hspace{.5truecm}0.59 \\
130 & \hspace{.5truecm}0.44 & \hspace{.5truecm}1.22 
\end{tabular} 
\end{center}

\noindent while $\delta\lambda_t /\lambda_t$ at $E_{CM}=1$~TeV and
1000 fb$^{-1}$ is

\begin{center}
\begin{tabular}{ccc}
$m_H(\mbox{GeV})$ & \hspace{.5truecm}$\epsilon_b\!=\!1$ & \hspace{.5truecm}$\epsilon_b\!=\!0.6$ \\
100 & \hspace{.5truecm}0.06 & \hspace{.5truecm}0.17 \\
110 & \hspace{.5truecm}0.06 & \hspace{.5truecm}0.18 \\
120 & \hspace{.5truecm}0.07 & \hspace{.5truecm}0.19 \\
130 & \hspace{.5truecm}0.08 & \hspace{.5truecm}0.22
\end{tabular} 
\end{center}

\noindent Our results qualitatively agree with those presented in
Ref.~\cite{merino}, if we assume $\epsilon_b\!=\!1$. A high efficiency
$b$-tag, along with of order $1000$ fb$^{-1}$ of integrated luminosity
will be critical for the top Yukawa coupling measurement.

%
\acknowledgments
The work of H.B and L.R. was supported in part by the U. S. Department
of Energy under contract number DE-FG02-97ER41022.  The work of
S.D. was supported by the U. S. Department of Energy under Contract
No. DE-AC02-76CH00016.  We are grateful to Frank Paige and Horst Wahl
for valuable discussions.
\newpage
%
\begin{table}
\begin{center}
\caption[]{The $\ell+4b+jets+\eslt$ final state at $E_{CM}\!=\!500$~GeV:
number of events for different selection cuts, assuming
$10^3\,$fb$^{-1}$ of integrated luminosity and $\epsilon_b\!=\!1$.  All
energies and masses are in GeV.  $m(rec)$ is the invariant mass of the
$b \bb$ system remaining after reconstructing the top quark
masses. $m(ex)$ is the invariant mass of the $b\bb$ system coming from
the Higgs decay artificially selected using Monte Carlo information.
$E_b$ is the energy of the slowest $b$-jet.}
\label{tab:eve_l_500}
\bigskip
\begin{tabular}{|c|c|c|c|c|c|c|}
\hline
channel & $H(100)$ & $H(110)$ & $H(120)$ & $H(130)$ & $t\tb
b\bb\ (EW)$ & $t\tb b\bb\ (QCD)$ \\
\hline
total & $960$ & $540$ & $250$ & $80$ & $170$ & $840$ \\
$\ell+4b+jets+\eslt$ & $33.6$ & $19.3$ & $9.8$ & $3.2$ & $6.7$ & $4.6$ \\
$E_b>25$ ; $m(rec.)>50$ & --- & $13.7$ & --- & --- & $4.2$ & $0.9$ \\
$E_b>40$ ; $m(rec.)>90$ & --- & --- & $2.2$ & --- & $0.35$ & $0.05$ \\
$E_b>45$ ; $m(rec.)>90$ & --- & --- & --- & $0.7$ & $0.15$ & $0.034$ \\
$E_b>25$ ; $m(ex.)>85$ & --- & $11.2$ & --- & --- & $1.6$ & $0.13$ \\
$E_b>40$ ; $m(ex.)>90$ & --- & --- & $3.5$ & --- & $0.3$ & $0.017$ \\
$E_b>45$ ; $m(ex.)>95$ & --- & --- & --- & $1.0$ & $0.034$ & $0.017$ \\
\hline
\end{tabular}
\end{center}
\end{table}
\begin{table}
\begin{center}
\caption[]{The $\ell+4b+jets+\eslt$ channel at $E_{CM}\!=\!500$~GeV:
estimated error on the top quark Yukawa coupling
($\delta\lambda_t/\lambda_t$) assuming $10^3\,$fb$^{-1}$ of integrated
luminosity and $\epsilon_b\!=\!1$ ($\epsilon_b\!=\!0.6$).}
\label{tab:err_l_500}
\bigskip
\begin{tabular}{|c|c|c|c|c|}
\hline
channel & $H(100)$ & $H(110)$ & $H(120)$ & $H(130)$ \\
\hline
$\ell+4b+jets+\eslt$ & $0.11\,\,(0.31)$ & $0.17\,\,(0.46)$ & $0.29\,\,(0.80)$ 
& $0.79\,\,(2.2)$ \\
$E_b>25$ ; $m(rec.)>50$ & --- & $0.18\,\,(0.49)$ & --- & --- \\
$E_b>40$ ; $m(rec.)>90$ & --- & --- & $0.39\,\,(1.09)$ & --- \\
$E_b>45$ ; $m(rec.)>90$ & --- & --- & --- & $0.74\,\,(2.05)$ \\
$E_b>25$ ; $m(ex.)>85$ & --- & $0.17\,\,(0.47)$ & --- & --- \\
$E_b>40$ ; $m(ex.)>90$ & --- & --- & $0.29\,\,(0.80)$ & --- \\
$E_b>45$ ; $m(ex.)>95$ & --- & --- & --- & $0.52\,\,(1.46)$ \\
\end{tabular}
\end{center}
\end{table}
\begin{table}
\begin{center}
\caption[]{The $4b\,+\ge 4\,jets\,$ final state at $E_{CM}\!=\!500$~GeV:
number of events for different selection cuts, assuming
$10^3\,$fb$^{-1}$ of integrated luminosity and $\epsilon_b\!=\!1$.}
\label{tab:eve_h_500}
\bigskip
\begin{tabular}{|c|c|c|c|c|c|c|}
\hline
channel & $H(100)$ & $H(110)$ & $H(120)$ & $H(130)$ & $t\tb b\bb\ (EW)$ & 
$t\tb b\bb\ (QCD)$ \\
\hline
total & $960$ & $540$ & $250$ & $80$ & $170$ & $840$ \\
$4b+\ge 4-jets$ & $32.8$ & $19.7$ & $8.8$ & $3.2$ & $5.8$ & $4.7$ \\
$E_b>25$ ; $m(rec.)>0$ & --- & $15.9$ & --- & --- & $4.4$ & $2.3$ \\
$E_b>40$ ; $m(rec.)>80$ & --- & --- & $2.0$ & --- & $0.89$ & $0.12$ \\
$E_b>45$ ; $m(rec.)>90$ & --- & --- & --- & $0.62$ & $0.3$ & $0.017$ \\
$E_b>25$ ; $m(ex.)>75$ & --- & $13.2$ & --- & --- & $3.3$ & $0.29$ \\
$E_b>40$ ; $m(ex.)>90$ & --- & --- & $2.6$ & --- & $0.42$ & $0.017$ \\
$E_b>45$ ; $m(ex.)>90$ & --- & --- & --- & $0.86$ & $0.31$ & $<0.017$ \\
\hline
\end{tabular}
\end{center}
\end{table}
\begin{table}
\begin{center}
\caption[]{The $4b\,+\ge 4\,jets\,$ channel at $E_{CM}\!=\!500$~GeV:
estimated error on the top quark Yukawa coupling
($\delta\lambda_t/\lambda_t$) assuming $10^3\,$fb$^{-1}$ of integrated
luminosity and $\epsilon_b\!=\!1$ ($\epsilon_b\!=\!0.6$).}
\label{tab:err_h_500}
\bigskip
\begin{tabular}{|c|c|c|c|c|}
\hline
channel & $H(100)$ & $H(110)$ & $H(120)$ & $H(130)$ \\
\hline
$4b+\ge 4-jets$ & $0.11\,\,(0.31)$ & $0.16\,\,(0.45)$ & $0.31\,\,(0.86)$ & 
$0.77\,\,(2.13)$ \\
$E_b>25$ ; $m(rec.)>50$ & --- & $0.17\,\,(0.47)$ & --- & --- \\
$E_b>40$ ; $m(rec.)>90$ & --- & --- & $0.50\,\,(1.39)$ & --- \\
$E_b>45$ ; $m(rec.)>90$ & --- & --- & --- & $0.90\,\,(2.50)$ \\
$E_b>25$ ; $m(ex.)>85$ & --- & $0.17\,\,(0.47)$ & --- & --- \\
$E_b>40$ ; $m(ex.)>90$ & --- & --- & $0.36\,\,(0.99)$ & --- \\
$E_b>45$ ; $m(ex.)>95$ & --- & --- & --- & $0.71\,\,(1.96)$ \\
\hline
\end{tabular}
\end{center}
\end{table}
\begin{table}
\begin{center}
\caption[]{The $\ell+4b+jets+\eslt$ final state at $E_{CM}\!=\!1$~TeV:
number of events for different selection cuts, assuming
$10^3\,$fb$^{-1}$ of integrated luminosity and $\epsilon_b\!=\!1$.}
\label{tab:eve_l_1000}
\bigskip
\begin{tabular}{|c|c|c|c|c|c|c|}
\hline
channel & $H(100)$ & $H(110)$ & $H(120)$ & $H(130)$ & $t\tb
b\bb\ (EW)$ & $t\tb b\bb\ (QCD)$ \\
\hline
total & $2420$ & $2080$ & $1690$ & $1210$ & $510$ & $1900$ \\
$\ell+4b+jets+\eslt $ & $111$ & $97$ & $81$ & $60$ & $29$ & $32$ \\
$m(rec.)>60$ & $104$ & --- & --- & --- & $28$ & $24$ \\
$m(rec.)>90$ & --- & $77$ & $68$ & --- & $17$ & $19$ \\
$m(rec.)>100$ & --- & --- & --- & $47$ & $15$ & $18$ \\
$m(ex.)>70$ & $97$ & --- & --- & --- & $25$ & $14$ \\
$m(ex.)>90$ & --- & $68$ & $67$ & --- & $7$ & $10$ \\
$m(ex.)>95$ & --- & --- & --- & $49$ & $4$ & $9$ \\
\hline
\end{tabular}
\end{center}
\end{table}
\begin{table}
\begin{center}
\caption[]{The $\ell+4b+jets+\eslt$ channel at $E_{CM}\!=\!1$~TeV:
estimated error on the top quark Yukawa coupling
($\delta\lambda_t/\lambda_t$) assuming $10^3\,$fb$^{-1}$ of integrated
luminosity and $\epsilon_b\!=\!1$ ($\epsilon_b\!=\!0.6$).}
\label{tab:err_l_1000}
\bigskip
\begin{tabular}{|c|c|c|c|c|}
\hline
channel & $H(100)$ & $H(110)$ & $H(120)$ & $H(130)$ \\
\hline
$\ell+4b+jets+\eslt $ & $0.07\,\,(0.19)$ & $0.08\,\,(0.21)$ & $0.09\,\,(0.24)$
 & $0.11\,\,(0.31)$ \\
$m(rec.)>60$ & $0.07\,\,(0.19)$ & --- & --- & --- \\
$m(rec.)>90$ & --- & $0.08\,\,(0.22)$ & $0.09\,\,(0.24)$ & --- \\
$m(rec.)>100$ & --- & --- & --- & $0.11\,\,(0.31)$ \\
$m(ex.)>70$ & $0.07\,\,(0.19)$ & --- & --- & --- \\
$m(ex.)>90$ & --- & $0.07\,\,(0.21)$ & $0.07\,\,(0.21)$ & --- \\
$m(ex.)>95$ & --- & --- & --- & $0.09\,\,(0.24)$ \\
\hline
\end{tabular}
\end{center}
\end{table}
\begin{table}
\begin{center}
\caption[]{The $4b\,+\ge 4\,jets\,$ final state at $E_{CM}\!=\!1$~TeV:
number of events for different selection cuts, assuming
$10^3\,$fb$^{-1}$ of integrated luminosity and $\epsilon_b\!=\!1$.}
\label{tab:eve_h_1000}
\bigskip
\begin{tabular}{|c|c|c|c|c|c|c|}
\hline
channel & $H(100)$ & $H(110)$ & $H(120)$ & $H(130)$ & $t\tb
b\bb\ (EW)$ & $t\tb b\bb\ (QCD)$ \\
\hline
total & $2420$ & $2080$ & $1690$ & $1210$ & $510$ & $1900$ \\
$4b+\ge 4-jets$ & $51$ & $48$ & $38$ & $27$ & $12$ & $19$ \\
$m(rec.)>65$ & $43$ & --- & --- & --- & $11$ & $12$ \\
$m(rec.)>75$ & --- & $39$ & --- & --- & $9$ & $11$ \\
$m(rec.)>90$ & --- & --- & $27$ & --- & $6$ & $9$ \\
$m(rec.)>95$ & --- & --- & --- & $19$ & $5$ & $9$ \\
$m(ex.)>65$ & $46$ & --- & --- & --- & $11$ & $8$ \\
$m(ex.)>75$ & --- & $41$ & --- & --- & $9$ & $6$ \\
$m(ex.)>90$ & --- & --- & $28$ & --- & $2$ & $5$ \\
$m(ex.)>95$ & --- & --- & --- & $21$ & $2$ & $4$ \\
\hline
\end{tabular}
\end{center}
\end{table}
\begin{table}
\begin{center}
\caption[]{The $4b\,+\ge 4\,jets\,$ channel at $E_{CM}\!=\!1$~TeV:
estimated error on the top quark Yukawa coupling
($\delta\lambda_t/\lambda_t$) assuming $10^3\,$fb$^{-1}$ of integrated
luminosity and $\epsilon_b\!=\!1$ ($\epsilon_b\!=\!0.6$).}
\label{tab:err_h_1000}
\bigskip
\begin{tabular}{|c|c|c|c|c|}
\hline
channel & $H(100)$ & $H(110)$ & $H(120)$ & $H(130)$ \\
\hline
$4b+\ge 4-jets$ & $0.10\,\,(0.29)$ & $0.11\,\,(0.30)$ & $0.13\,\,(0.36)$ & 
$0.17\,\,(0.48)$ \\
$m(rec.)>65$ & $0.11\,\,(0.30)$ & --- & --- & --- \\
$m(rec.)>75$ & --- & $0.11\,\,(0.32)$ & --- & --- \\
$m(rec.)>90$ & --- & --- & $0.14\,\,(0.39)$ & --- \\
$m(rec.)>95$ & --- & --- & --- & $0.18\,\,(0.50)$ \\
$m(ex.)>65$ & $0.10\,\,(0.28)$ & --- & --- & --- \\
$m(ex.)>75$ & --- & $0.10\,\,(0.28)$ & --- & --- \\
$m(ex.)>90$ & --- & --- & $0.11\,\,(0.32)$ & --- \\
$m(ex.)>95$ & --- & --- & --- & $0.14\,\,(0.38)$ \\
\hline
\end{tabular}
\end{center}
\end{table}
%
   
%
%
%

\newpage
\iftightenlines\else\newpage\fi
\iftightenlines\global\firstfigfalse\fi
\def\dofig#1#2{\epsfxsize=#1\centerline{\epsfbox{#2}}}
%
\begin{figure}
\dofig{4.5in}{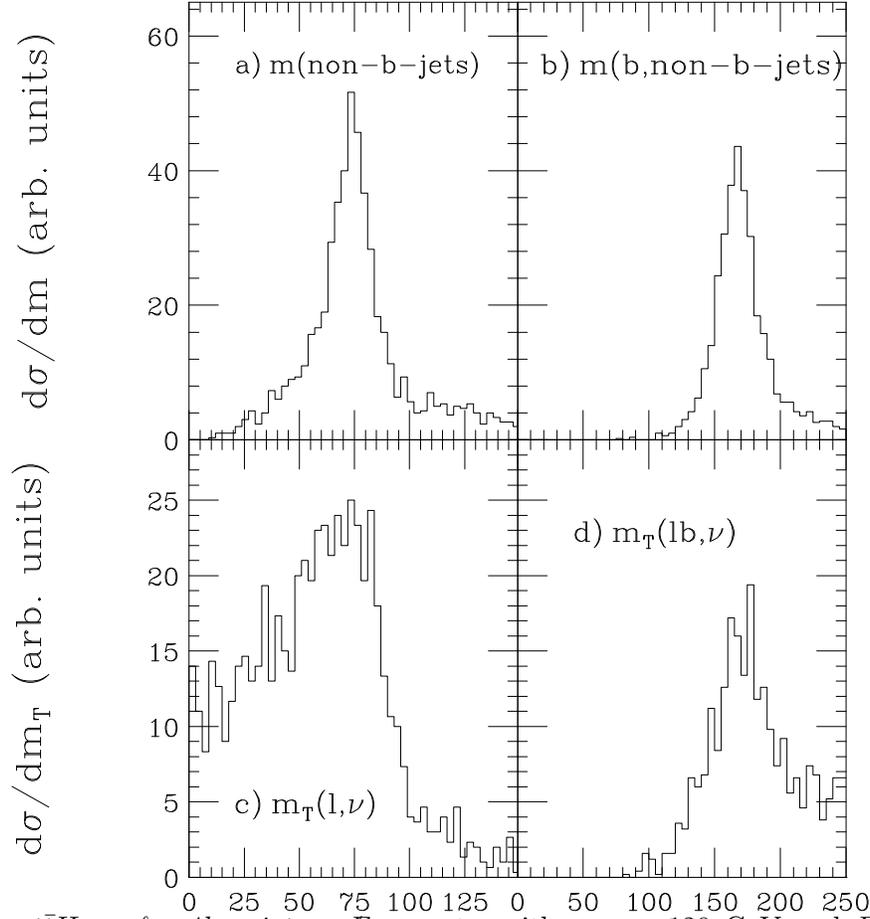}
\caption[]{For $t\tb H\to\ell +4b+jets+\eslt$ events, 
  with $m_H=120$~GeV and $E_{CM}=1$~TeV, we plot distributions in
  {\it a}) invariant mass of non-$b$-jets, {\it b}) invariant mass of
  non-$b$-jets plus the $b$-jet which gives a mass closest to $m_t$,
  {\it c}) isolated lepton plus missing energy transverse mass, and
  {\it d}) isolated lepton plus $b$-jet plus missing energy cluster
  transverse mass for the $b$-jet which most closely reconstructs
  $m_t$.}
\label{fig:masses}
\end{figure}
%
%
%
\begin{figure}
\dofig{5in}{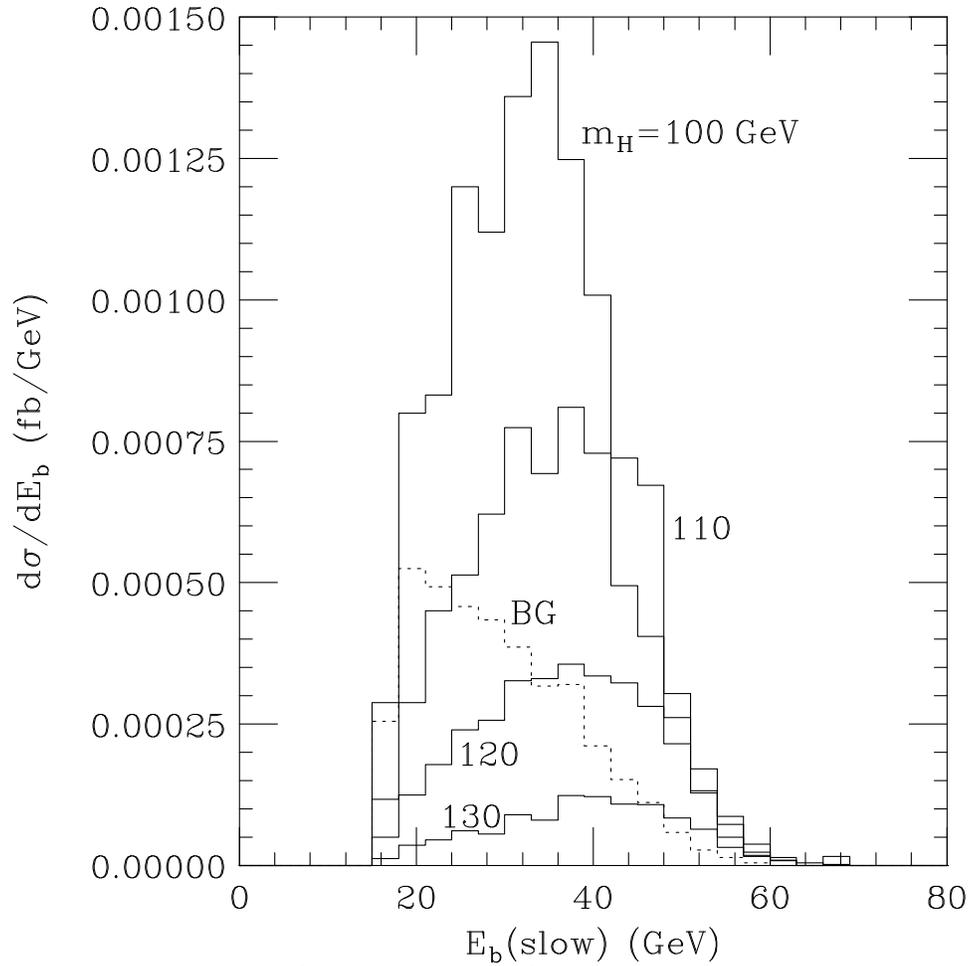}
\caption[]{
  Energy distribution for the slowest of four $b$-jets in $\ell
  +4b+jets+\eslt$ events at the NLC, at $E_{CM}=500$~GeV.}
\label{fig:ebsl}
\end{figure}
%
\begin{figure}
\dofig{5in}{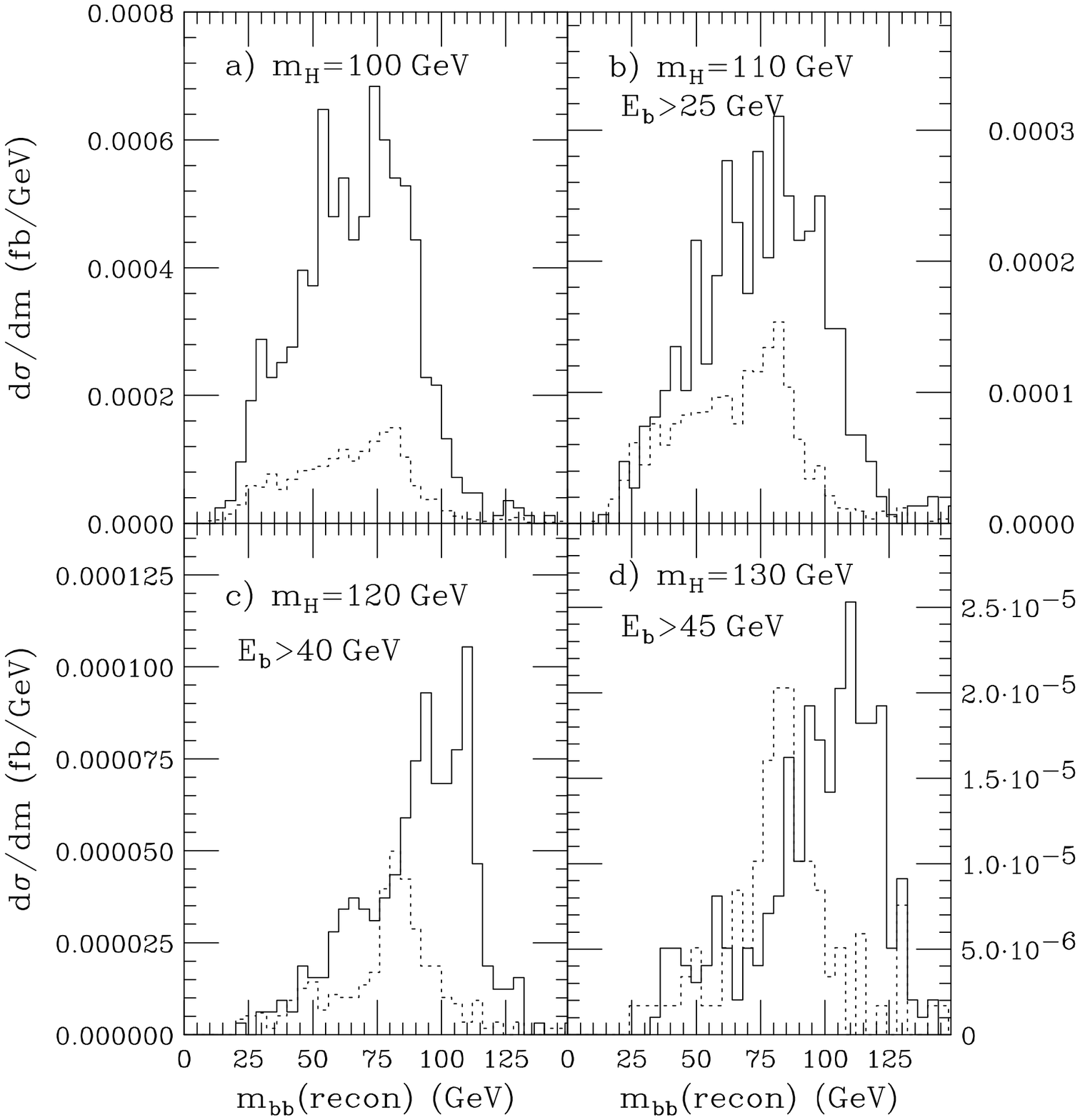}
\caption[]{
  Distribution in $b\bb$ invariant mass for the two remaining $b$-jets
  after top mass reconstruction, for the semileptonic events at
  $E_{CM}=500$~GeV.  Signal is solid, while the sum of EW and QCD
  background is dashed.}
\label{fig:xmbbl5}
\end{figure}
%
\begin{figure}
\dofig{5in}{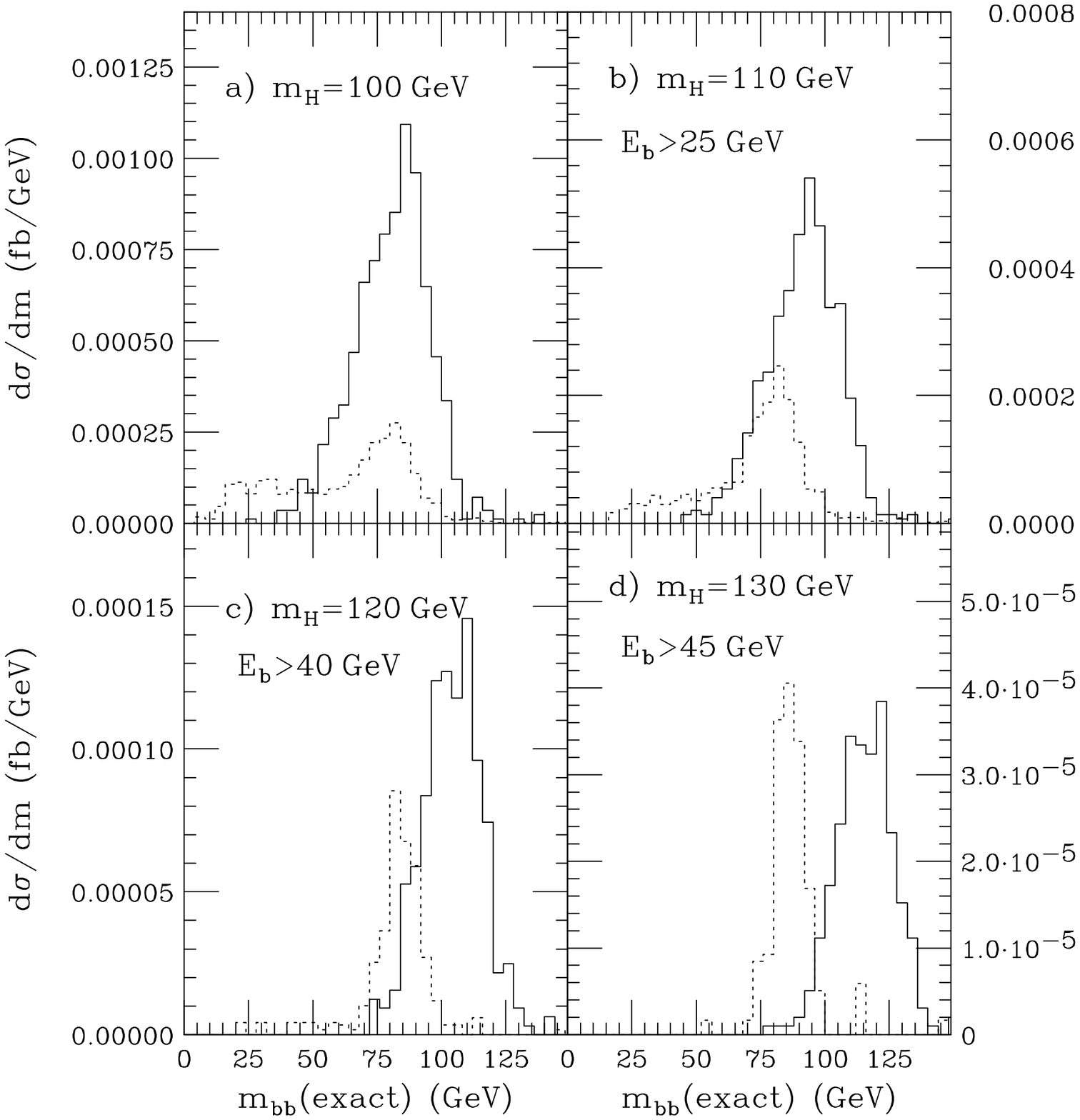}
\caption[]{
  Distribution in $b\bb$ invariant mass for the two correctly
  identified non-top $b$-jets, using generator information for
  semileptonic events, at $E_{CM}=500$~GeV.  Signal is solid, while
  the sum of EW and QCD background is dashed.}
\label{fig:xmbbexl5}
\end{figure}
%
\begin{figure}
\dofig{5in}{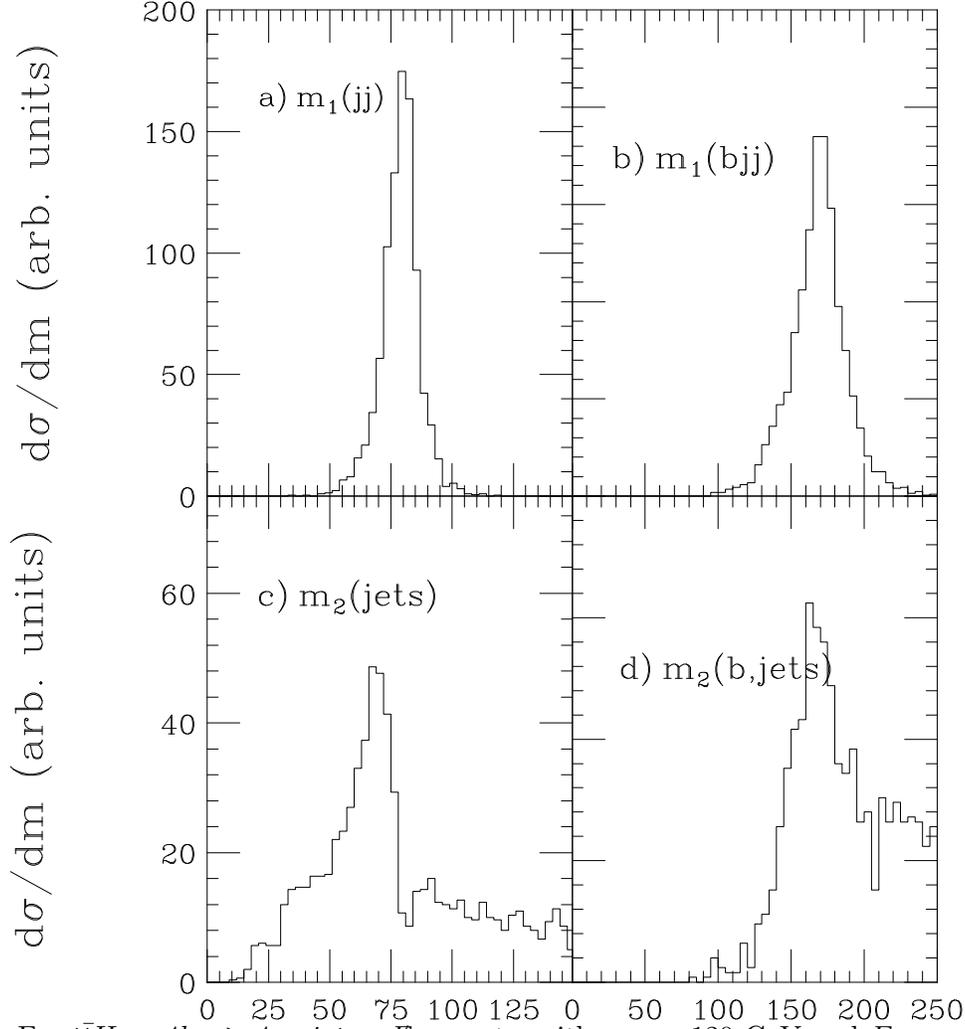}
\caption[]{For $t\tb H\to 4b+\ge 4-jets+\eslt$ events, 
  with $m_H=120$~GeV and $E_{CM}=1$~TeV, we plot distributions in
  {\it a}) dijet invariant mass closest to $M_W$, 
  {\it b}) invariant mass of dijet plus $b$-jet combination closest to $m_t$,
  {\it c}) invariant mass of remaining non-$b$-jets, and
  {\it d}) invariant mass of remaining non-$b$-jets plus remaining $b$-jet
  which gives a mass closest to $m_t$.}
\label{fig:massh}
\end{figure}
%
\begin{figure}
\dofig{5in}{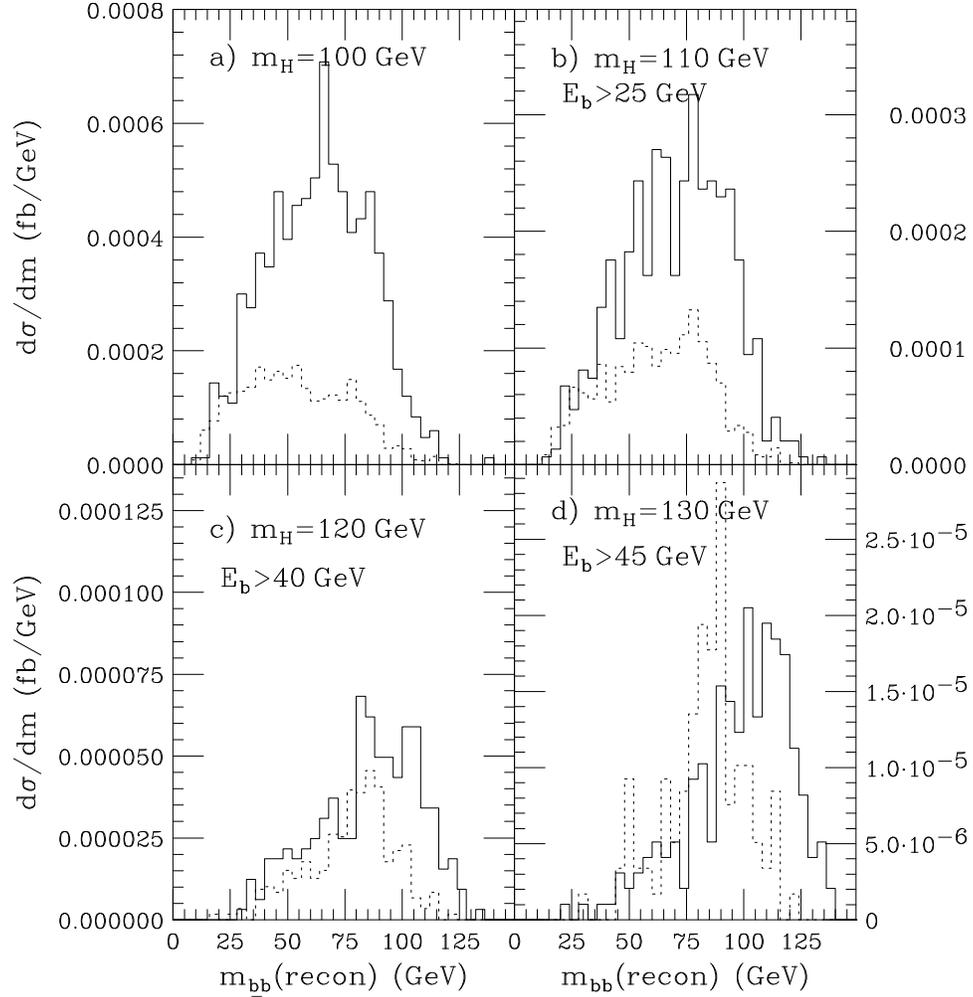}
\caption[]{
  Distribution in $b\bb$ invariant mass for the two remaining $b$-jets
  after top mass reconstruction at $E_{CM}=500$~GeV, for the
  hadronic final state.  Signal is solid, while the sum of EW and QCD
  background is dashed.}
\label{fig:xmbbh5}
\end{figure}
%
\begin{figure}
\dofig{5in}{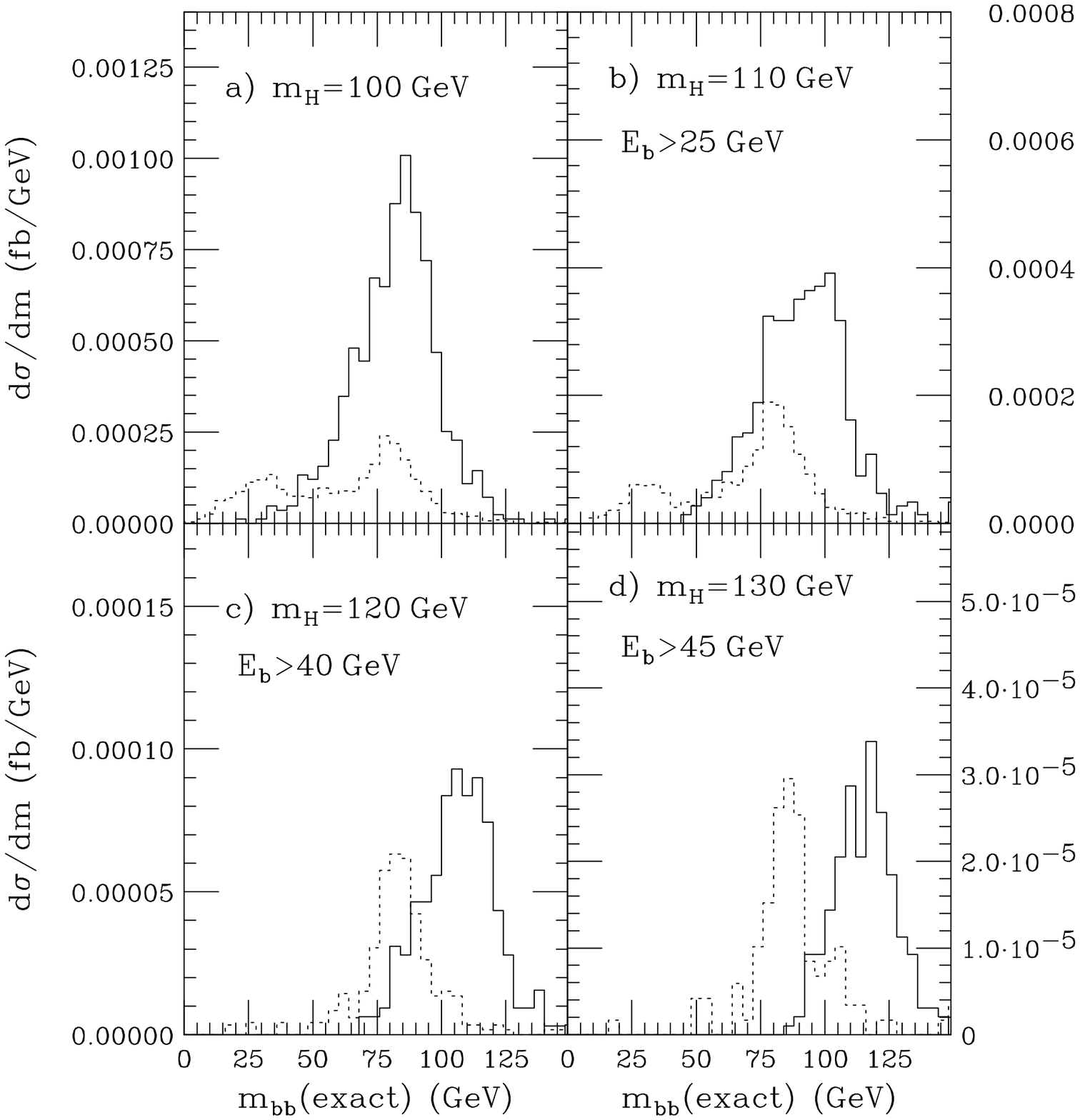}
\caption[]{
  Distribution in $b\bb$ invariant mass for the two correctly
  identified non-top $b$-jets, using generator information for
  hadronic events, at $E_{CM}=500$~GeV. Signal is solid, while the
  sum of EW and QCD background is dashed.}
\label{fig:xmbbexh5}
\end{figure}
%
\begin{figure}
\dofig{5in}{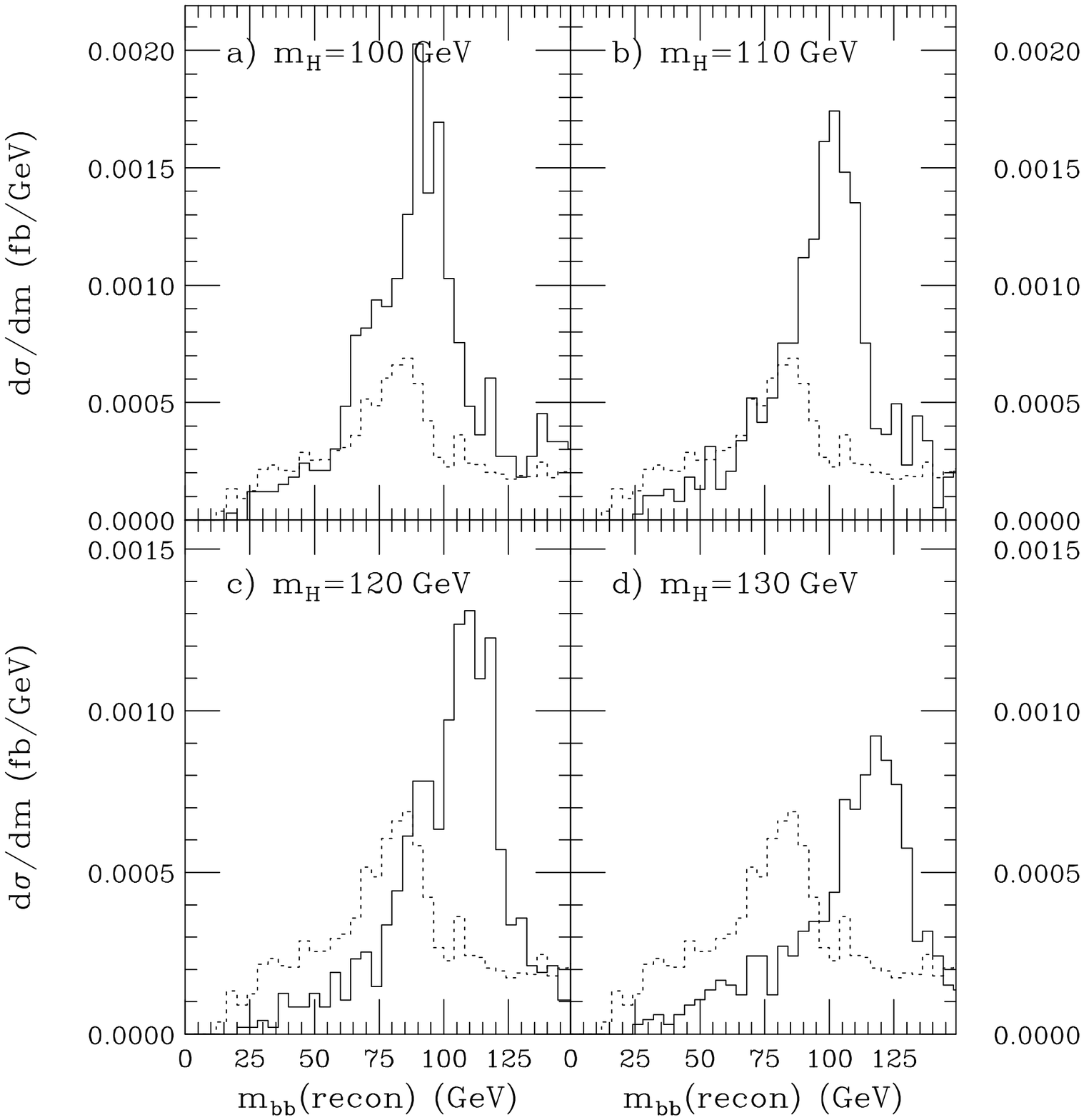}
\caption[]{
  Distribution in $b\bb$ invariant mass for the two remaining $b$-jets
  after top mass reconstruction, for semileptonic events at
  $E_{CM}=1$~TeV.  Signal is solid, while the sum of EW and QCD
  background is dashed.}
\label{fig:xmbbl1}
\end{figure}
%
\begin{figure}
\dofig{5in}{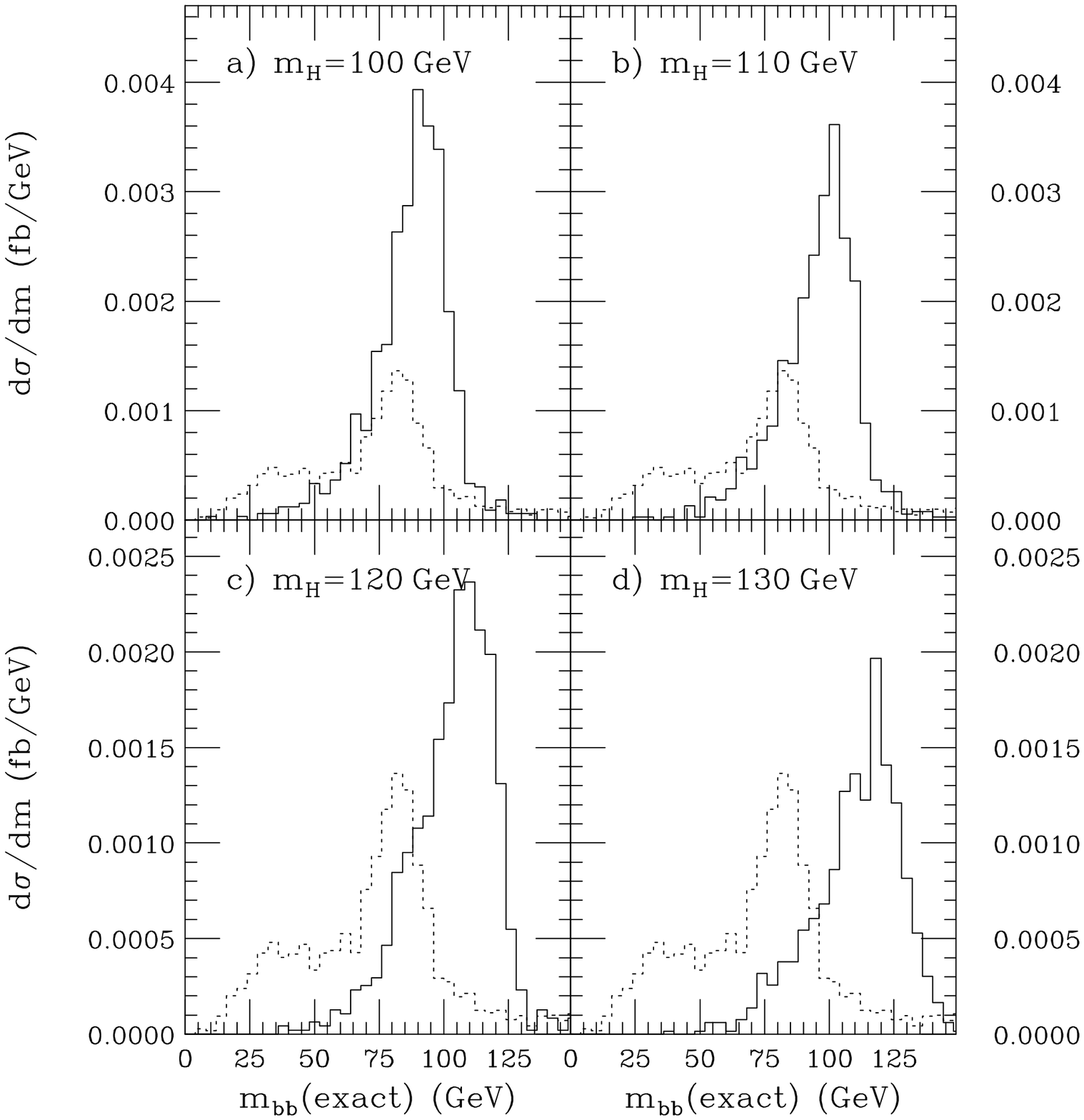}
\caption[]{
  Distribution in $b\bb$ invariant mass for the two correctly
  identified non-top $b$-jets, using generator information for
  semileptonic events, at $E_{CM}=1$~TeV.  Signal is solid, while
  the sum of EW and QCD background is dashed.}
\label{fig:xmbbexl1}
\end{figure}
%
\begin{figure}
\dofig{5in}{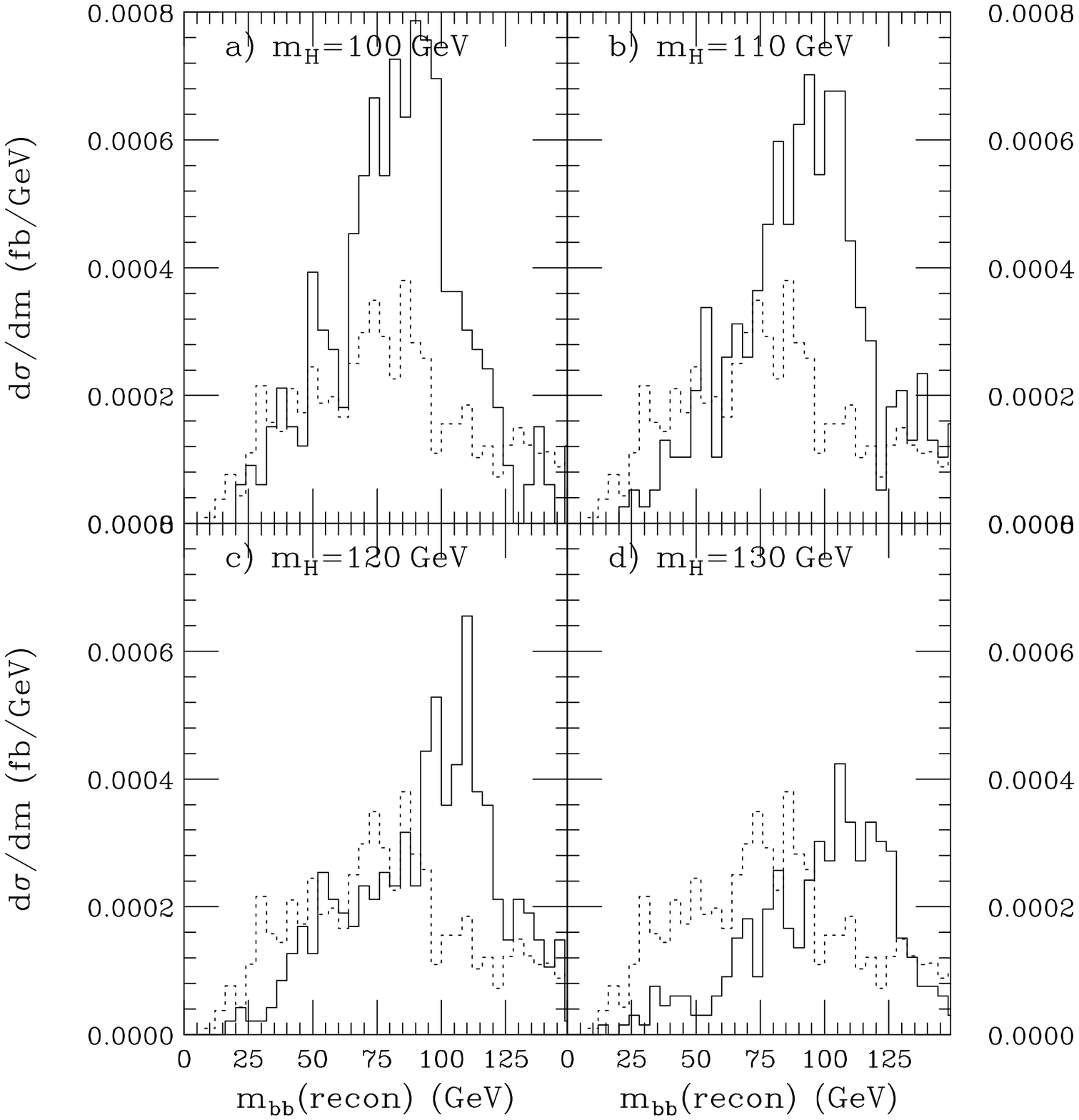}
\caption[]{
  Distribution in $b\bb$ invariant mass for the two remaining $b$-jets
  after top mass reconstruction in hadronic events at $E_{CM}=1$~TeV.
  Signal is solid, while the sum of EW and QCD background is dashed.}
\label{fig:xmbbh1}
\end{figure}
%
\begin{figure}
\dofig{5in}{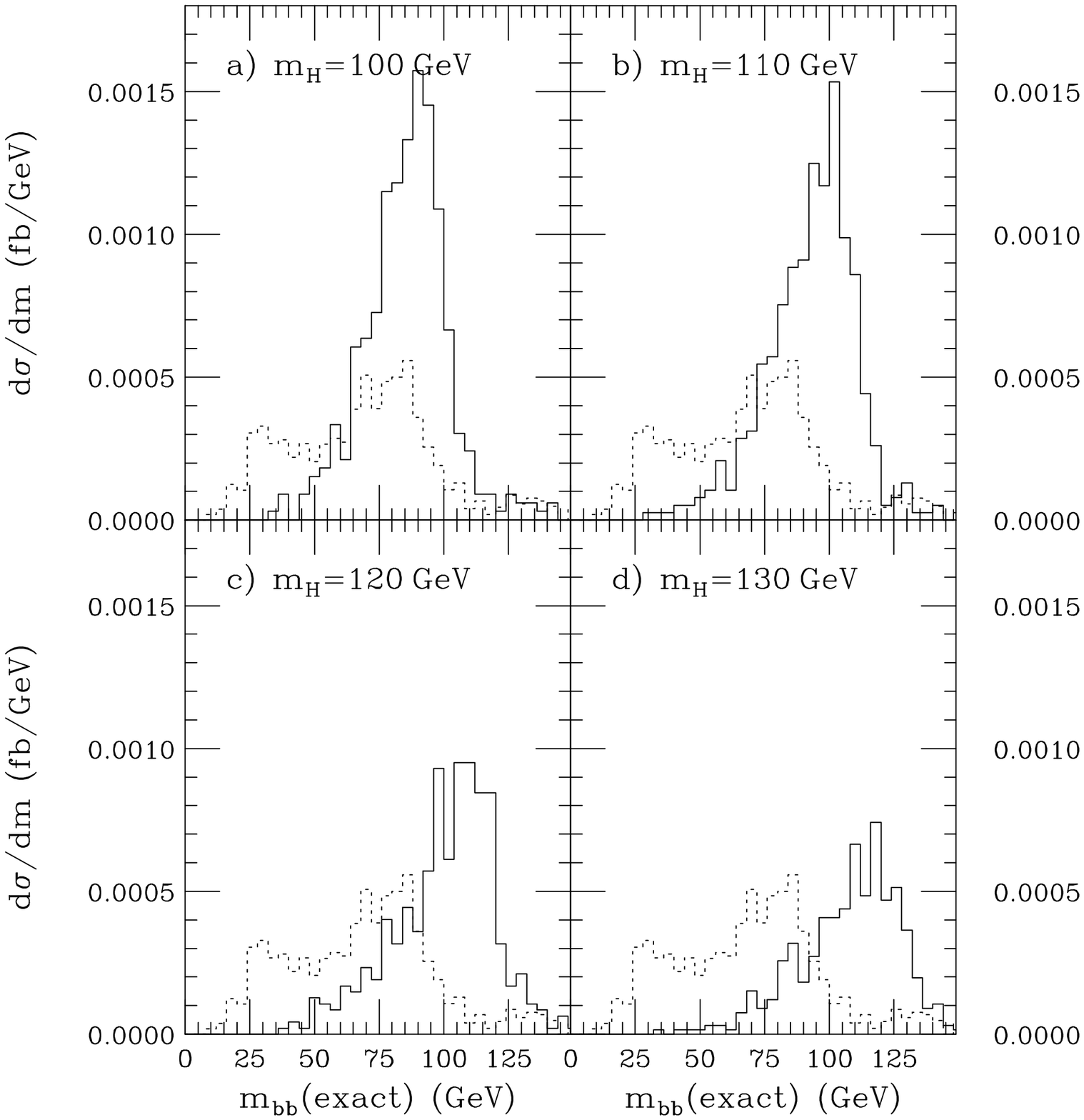}
\caption[]{
  Distribution in $b\bb$ invariant mass for the two correctly
  identified non-top $b$-jets, using generator information for
  hadronic events, for $E_{CM}=1$~TeV.  Signal is solid, while the
  sum of EW and QCD background is dashed.}
\label{fig:xmbbexh1}
\end{figure}
\end{document}